\def\year{2022}\relax
\documentclass[letterpaper, anonymous=false]{article} 
\usepackage{aaai24}  
\usepackage{times}  
\usepackage{helvet}  
\usepackage{courier}  
\usepackage[hyphens]{url}  
\usepackage{graphicx} 
\usepackage{natbib}  
\usepackage{caption} 
\DeclareCaptionStyle{ruled}{labelfont=normalfont,labelsep=colon,strut=off} 
\usepackage{algorithm}
\usepackage{algorithmic}

\usepackage{newfloat}
\usepackage{listings}

\usepackage{multirow}
\usepackage{xspace}
\usepackage{xcolor}

\floatstyle{ruled}
\newfloat{listing}{tb}{lst}{}
\floatname{listing}{Listing}

\newif\ifstatus
\statusfalse

\newcommand\jose[1]{\ifstatus\textcolor{red}{#1 --- js}\fi}
\newcommand\vahid[1]{\ifstatus\textcolor{violet}{#1 --- vg}\fi}

\title{A Holistic Indicator of Polarization to Measure Online Sexism}

\author{
Vahid Ghafouri,\textsuperscript{\rm 1,2}
Jose Such,\textsuperscript{\rm 3,4}
Guillermo Suarez-Tangil,\textsuperscript{\rm 2}
}
\affiliations {
\textsuperscript{\rm 1}IMDEA Networks Institute,\\
\textsuperscript{\rm 2}Universidad Carlos III de Madrid,\\
\textsuperscript{\rm 3}King's College London,\\
\textsuperscript{\rm 4}VRAIN, Universitat Politecnica de Valencia,\\
\{vahid.ghafouri, guillermo.suarez-tangil\}@imdea.org, jose.such@kcl.ac.uk

}

\usepackage{bibentry}

\begin{document}

\maketitle

\begin{abstract}
The online trend of the manosphere and feminist discourse on social networks requires a holistic measure of the level of sexism in an online community. This indicator is important for policymakers and moderators of online communities (e.g., subreddits) and computational social scientists, either to revise moderation strategies based on the degree of sexism or to match and compare the temporal sexism across different platforms and communities with real-time events and infer social scientific insights.

In this paper, we build a model that can provide a comparable holistic indicator of toxicity targeted toward male and female identity and male and female individuals. Despite previous supervised NLP methods that require annotation of toxic comments at the target level (e.g. annotating comments that are specifically toxic toward women) to detect targeted toxic comments, our indicator uses supervised NLP to detect the presence of toxicity and unsupervised word embedding association test to detect the target automatically.

We apply our model to gender discourse communities (e.g., r/TheRedPill, r/MGTOW, r/FemaleDatingStrategy) to detect the level of toxicity toward genders (i.e., sexism). 
Our results show that our framework accurately and consistently (93\% correlation) measures the level of sexism in a community. 
We finally discuss how our framework can be generalized in the future to measure qualities other than toxicity (e.g. sentiment, humor) toward general-purpose targets and turn into an indicator of different sorts of polarizations.
\end{abstract}

\maketitle

\section{Introduction}
\label{sec:introduction}

Polarization and radicalization of opinion on social media have been a hot topic of research in the recent Computational Social Science literature \cite{Bail2018}. 
One type of polarization on social media can be based on people's views about gender roles and identity which can be partially observed by looking into the use of biased language on different sides of the online gender discourse spectrum. For instance, 
prior work has studied the use of toxic and misogynistic language in manosphere  (e.g., r/TRP, r/MGTOW) communities on social networks~\cite{Dignam2019,Ging2019,Farrel2019,icwsm21manosphere}. 
However, there is a wide gap in both qualitative and quantitative studies offering a measure that can precisely \emph{quantify} the \emph{level} of sexism inside every community at scale. 
In other words, previous research tells us that community A is sexist, but it doesn't say ``{\em how much exactly.}''; or ``{\em between community A and B which one is more sexist?}''



In this paper, we define a macroscopic scalar indicator that can give us an overall measure of the total toxicity aimed toward male and female identity in a community (in our case-study, subreddits). Our scalar indicator is based on the combination of three parameters for each adjective inside a community where each parameter preserves one of the following key qualities of our work models: 1) How toxic is a word's context within a community's discourse? 2) How frequently it has been used inside its corpus? 3) How biased is the word toward a gender in that community?  

The first parameter is based on a supervised NLP model that detects whether a sentence is toxic or not; without the need to judge the target of the toxicity. Then it computes the rate of a word's appearance in a toxic sentence to calculate the toxicity of a word's context. This is more reasonable than previous works that solely look into the polarization \cite{ferrer2021discovering} or toxicity \cite{Farrel2019} of words using a dictionary of polarized or toxic lexicons as a word can appear less or more toxic in different discourses.

Existing methods suffer limitations in identifying the target towards which toxicity is directed as when it comes to annotation, the toxic comments towards a very specific group identity are sparse. Measuring targeted toxicity toward various group-identities in a fully supervised manner requires a separate manual annotation of comments that are specifically toxic toward each group. One of our contributions is to keep the supervision in the first parameter, to merely decide the toxicity rate, and introduce the third parameter, which is unsupervised, to measure the target of the toxicity automatically.

The third parameter is based on the idea of Word Embedding Association Test (WEAT) \cite{Caliskan2017} that defines the gender bias of a word by looking into its word-embedding cosine similarity with embeddings of gender-related words (e.g. woman, she, female), namely ``attribute sets''. However, it makes no distinction between bias toward gender identity and individual characters from a gender. Meaning that there are cases where individual female characters, like a female politician, are targeted rather than all women as a group-identity. 
This obfuscates the quantification of gender bias, making metrics indicating the level of bias coarse-grained and ineffective at distinguishing other underlying motives (i.e. political motives). 
Thus, attributing an adjective to several female characters using existing works is computationally equivalent to attributing it to women in general, since most works based on word-embedding associations, mix both gender identity terms (e.g. men, women) and gender pronouns (e.g. he, she) in their attribute sets. 
By separating the two, we define two complementary indicators; one indicating the toxicity toward male/female identity, and one measuring the toxicity toward individual male/female figures.
These two indicators are more informative on their own than when they are aggregated.


In summary, we make the following contributions: 
\begin{itemize}
    \item We propose a model that can measure \emph{various sorts of polarization} on social networks through a \emph{scalar value} that can be used to compare disparate communities.
    \item We offer a clear distinction between toxicity targeted toward gender \emph{identity} and toxicity targeted toward \emph{individual} male and female characters and we quantify each of them separately. 
    \item We calculate the toxicity of words based on their context in a corpus to address the limitations of previous context-unaware lexicon-based approaches \cite{Hine2017,Fast2016,Farrel2019}. Finally, we apply a unique holistic model to several subreddits from various sides of the gender-discourse spectrum and report the targeted toxicity level for each.
\end{itemize}

\section{Related Work}
\label{sec:related-work}

Since our sexism indicator combines the notion of unsupervised word-embedding associations 
with a supervised toxic comments classification, we divide our literature review into two subsections. In the first part, we discuss the previous works which have tried to quantify language bias based on word-embeddings, and in the second subsection, we review some previous efforts on toxic comment classification.

\subsection{Language Bias Quantification Based on Word-Embeddings}
\label{subsec:LBQboWE}


In \citeyear{Greenwald1998}, \citeauthor{Greenwald1998} developed Implicit Association Test (IAT) as an experimental method for identifying such implicit biases for every user \cite{Greenwald1998}. The test tends to measure the strength of implicit associations between attribute concepts (e.g., black people, or LGBTQ+ members) and evaluations (e.g., good, or bad) or stereotypes (e.g., athletic, or clumsy) based on the time it takes for a user to assign each word to the attribute concepts.

Inspired by IAT in clinical psychology, Caliskan et al. \cite{Caliskan2017} leveraged the emergence of word-embeddings in NLP, to develop the Word Embedding Association Test (WEAT) in order to confirm the existence of similar implicit/explicit associations; based on the relative distances of attribute words vectors with target concepts' word vectors. For instance, WEAT shows that science-related terms' vectors are closer to the word \(\vec{man}\) than the word \(\vec{woman}\), in contrast with art-related terms which have more cosine similarity with the word \(\vec{woman}\) than \(\vec{man}\). 

However, since the sets of target words (e.g., science, art) and attribute words (i.e., any dual concept, like men and women) in WEAT are determined by humans, the user can cherry-pick the set of terms to witness the desired outcome \cite{Ethayarajh2019}. 
This means that WEAT and subsequent works are more suitable when the researcher is aware of a predefined set of biased concepts in real-world data (e.g., IAT test in clinical psychology research) and is trying to validate that those biases appear in a text-corpus. 

There is scarce prior work aiming at \emph{discovering} biases~\cite{Ghafouri2023kialo} in word-embeddings rather than \emph{confirming} them \cite{ferrer2021discovering}. 
\citeauthor{Swinger2018} makes some attempt in this direction \cite{Swinger2018}, however, their work still relies on crowd-sourcing and human judgment to assess if the biases in the word-embeddings match prevalent stereotypes in the real world. 
Moreover, prior work based on word-embedding associations merely focuses on quantifying gender biases in certain word sets (whether predefined or automated), yet does not offer a systematic way to compute an \emph{overall} measurement of gender bias and sexism that is comparable across communities.

\subsection{Toxic Comment Detection}
\label{subsec:toxiccommentdetection}

Related work has leveraged NLP to detect different types of toxic language such as aggression \cite{Aroyehun2018}, hate-speech~\cite{Burnap2015}, and offensive language~\cite{singh2024differences}. Moreover, IberLEF 2021\footnote{\url{https://sites.google.com/view/iberlef2021/}} has introduced EXIST,\footnote{\url{http://nlp.uned.es/exist2021/}} a hierarchical NLP classification task with an annotated dataset of \textit{sexist} vs \textit{non-sexist} tweets at level 1, and a categorization of the type of sexism (if applicable) at level 2 (e.g., stereotyping, objectification, sexual and non-sexual violence, etc.). 
Next to translation-augmentation methods, participants applied classical and Deep NLP on the task where pre-trained Deep NLP models (e.g., BERT) slightly outperformed the classical NLP methods \cite{Subies2021, Schtz2021, he2023prompt}.


The most relevant to our intention in this paper is OffensEval2019\footnote{\url{https://sites.google.com/site/offensevalsharedtask/offenseval2019}} task shared on SemEval2019\footnote{\url{https://alt.qcri.org/semeval2019/}} by \citeauthor{Zampieri2019offenseval} that looks into the \textit{type} and \textit{target} of the toxicity simultaneously. A hierarchical classification task consists of three sub-tasks: First, identification of the offensive language (i.e. \textit{offensive} or \textit{not}). The second sub-task would be to detect whether the offense is \textit{targeted} or \textit{not}, and the third is to check whether this targeted offense is targeted toward a \textit{group} or toward a \textit{person}. All these works use fully-supervised NLP. 
However, fully-supervised NLP toxicity detection tasks are highly prone to distribution-shift; an effect that happens where the distribution of the train-set is different from the test-set \cite{Koh2021}, causing the NLP model to yield lower accuracy on test-sets that are out of its sampled train data. Moreover, supervised NLP tasks are also prone to the \textit{concept-drift} effect. Concept-drift happens when ``algorithms trained on annotated data in the past may under-perform when applied to contemporary data'' \cite{Muller2020}. Therefore, reducing the level of supervision is an important agenda to follow. 

Generally, for a community analysis level, our framework provides a less supervised and more flexible solution for measuring targeted toxicity. Less supervised, because it only requires the data labeled in the first subtask (i.e., offensive/toxic or not), and the second and the third sub-tasks will be embedded in the effect of unsupervised word-embedding biases added to the model. More flexible, because the attribute words can be altered arbitrarily to detect other types of targeted toxicity, and the supervised data can change to detect qualities other than toxicity (e.g., polarity). To achieve this goal, we create a model that combines the existing notion of unsupervised word-embedding associations with novel semi-supervised tasks that build on recent efforts for toxic content detection.


\section{Methodology}
\label{sec:method}

\begin{figure*}[h]
  \centering
  \includegraphics[width=0.75\textwidth]{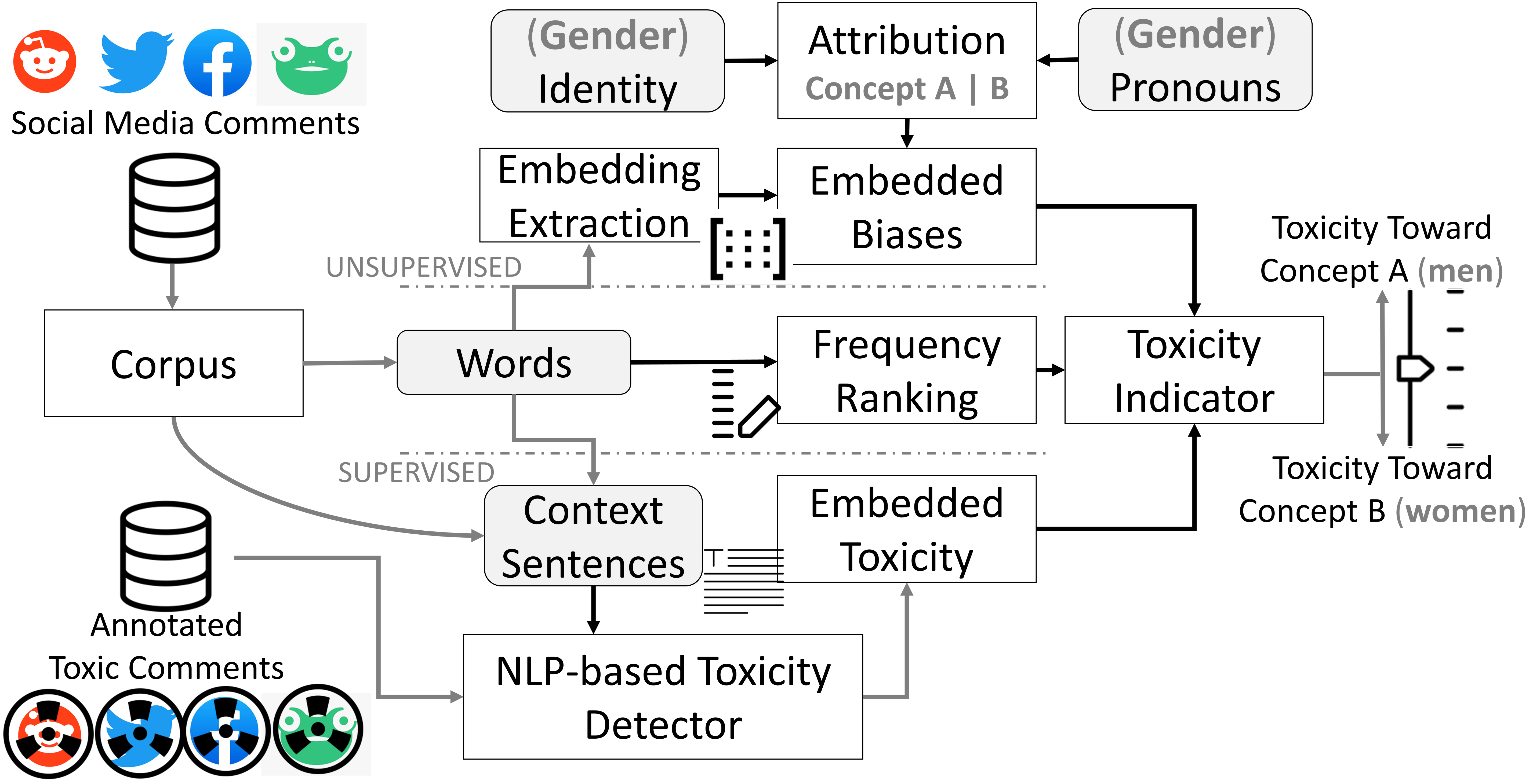}
  \caption{Outlook of our processing pipeline.}
  \label{fig:big-picture}
\end{figure*}

Figure \ref{fig:big-picture} shows an overview of our processing pipeline, which takes a corpus as input and returns an indicator of polarization as output. 
The area above the top gray dash-line shows the unsupervised nature of our work designed to measure the word-embedding biases.
The area below the bottom gray dash-line depicts the supervised pipeline we build to measure the toxicity embedded inside each adjective. The area between the two gray lines refers to the frequency-percentile ranking; another parameter that we take into account. 

\subsection{Preliminaries}
\label{subsec:preliminaries}
The sexism scalar indicator we provide consists of three variables: \textit{Embedded-Toxicity}, \textit{Frequency-Percentile-Ranking}, and \textit{Embedding-Bias} which we calculate for every adjective term in each corpus separately. In this section, we will explain in detail how we measure these three variables:

\subsubsection{\textbf{Embedded Toxicity}}

Every community has a set of terms and idioms that may preserve different meanings in the context of that community than their universal meanings \cite{ferrer2021discovering,garcia2018context}.
Looking at gender-related discourse, for instance, some terms that are considered \textit{neutral} when viewed out of context can actually carry negative sentiments and even toxicity. 
Words ``flirtatious'', and ``hypergamous'' can be used in manosphere discourse to manifest negative opinions about women's sexual lifestyle, or the word ``casual'' can be used to encourage only having casual sex with a group of people.
The term ``unicorn'', for example, is also a common term in gender-related communities (e.g. r/TheRedPill) to refer to unrealistic views about an ideal partner that could also be accompanied by toxic ideas around itself.

Thus, since computing words' toxicity is our objective in this section, it is vital to have a metric that computes a word's embedded toxicity according to its context, rather than dictionary-based analyses (e.g. Weaponized-Word\footnote{\url{https://weaponizedword.org/}} ) that pre-define a word's toxicity according to its global context \cite{paul2016anonymine,ireland2020profiling}.

\begin{figure*}[h]
  \centering
  \includegraphics[width=\linewidth]{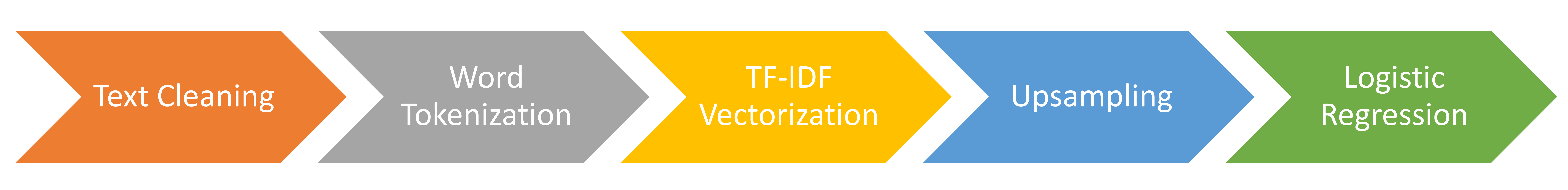}
  \caption{Processing pipeline for building our Toxicity-Detector NLP model.}
  \label{fig:nlp-pipline}
\end{figure*}

In order to calculate each word's embedded toxicity while covering the word's context, we propose a count-based semi-supervised method. First, we use our annotated toxic comments dataset (see Section~\ref{sec:datasets}) to build a Toxicity-Detector machine based on a supervised NLP classification model that can predict a comment as \textit{non-toxic (0)} or \textit{toxic (1)}. 
Building the model goes through the NLP pipeline as presented in Figure \ref{fig:nlp-pipline}. We initially clean and preprocess the text (e.g. removing stop-words, removing punctuation marks, lemmatization). Then, we tokenize the sentences and convert them into TF-IDF vectors. We add upsampling to balance the classes' size as the ``Toxic'' class is as 30\% big as the Non-Toxic class. Finally, we split the data into train (70\%) and test (30\%) sets and pass it through a Logistic Regression.
The F1-Score (macro) on the test-set is above 91\%. We also compare this accuracy with an advanced transformer-based model in Section~\ref{sec:evaluation} and show that our classic NLP model maintains a close performance to it while demanding a significantly lower computational cost.


After building the Toxicity-Detector model, we use it to systematically annotate every sentence in a community's data as \textit{Toxic} or \textit{Non-Toxic} (1 or 0). 
Then, for every adjective word that appears in the community, we average through the labels of the sentences that contain the adjective. We also ignore words belonging to other parts of speech as suggested by Ferrer et al \cite{ferrer2021discovering}.
Let \(sen_j\) represent sentence \(j\) in a corpus and \(ToxicityDetector()\) denote the function that calculates sentences' toxicity. Then, the \textit{Embedded-Toxicity} of word \(i\), \(T_{w_i}\), would be calculated by Equation \ref{equation:Toxicity}:

\begin{equation}
    T_{w_i} = \frac{\sum_{j}\{ToxicityDetector(sen_j) \mid sen_j\ni w_i\}}{|\{sen_j\mid sen_j\ni w_i\}|}
    \label{equation:Toxicity}
\end{equation}

Now assume, for instance, the word ``casual'' in r/TheRedPill which is neutral globally yet toxic locally. 
The Embedded-Toxicity parameter sees a sentence like ``You must only exploit her for casual sex and dump her'' labeled as toxic, and due to its context gives a higher score to the toxicity of the word ``casual''.

\subsubsection{\textbf{Frequency Ranking}}

The frequency of a word in a community's corpus is another important parameter that has to be taken into account in the final metric. Considering that we are studying toxicity as a scalable, community-wide metric, it makes sense to amplify the effect of the most frequent types of toxicity, over those that happen rarely.  

%
One option to preserve the effect of frequency in our metric is to simply weigh frequencies and biases. However, according to Zipf's law, the term-frequency gap inside a corpus increases exponentially as we move toward the top frequent words \cite{Newman2004}. 
This could cause the frequency to dominate other parameters and distort the balance we intend to preserve between bias, frequency, and embedded toxicity. 
Thus, we convert the raw frequency of adjectives into frequency-percentile-ranking to smooth the effect of the frequency in our model, i.e.: the percentage of the adjectives that an adjective outnumbers~\cite{ferrer2021tkde}. 
In addition to smoothing, it also creates a more scalable output as it provides a parameter between 0 and 1, that is compatible with our two other parameters.

Assuming that \(V\) represents the vocabulary of all adjectives in a corpus, and \(f_{w_i}\) denotes the frequency of word \(i\), the frequency-percentile-ranking of each word \(FPR_{w_i}\) would be calculated by Equation~\ref{equation:FPR}.

\begin{equation}
    FPR_{w_i} = \frac{|\{w_j\in V \mid f_{w_i}>f_{w_j}\}|}{|V|}
    \label{equation:FPR}
\end{equation}

\subsubsection{\textbf{Embedding-Bias}}
\label{sec:method:preliminaries:embeddingbias}

The final parameter is the embedding-bias which is supposed to measure the level of bias a word has toward a targeted concept. We follow the idea of the Word Embedding Association Test (WEAT) and several subsequent papers in quantifying global gender biases in words based on word embeddings \cite{Caliskan2017,Bolukbasi2016,ferrer2021discovering} and apply it to different communities' corpora to obtain the gender bias of every adjective in each corpus. This is to quantify how much an adjective in a corpus points its finger toward a certain group.

In this method, we take two sets of attribute words related to two distinct concepts (male and female in this case) and represent each set by the element-wise average of its word-embedding vectors.

Let \(S_{A}=\{{w_0, w_1, ... w_n}\}\) and \(S_{B}=\{{w_0, w_1, ... w_n}\}\) denote two sets of words that represent two different attribute concepts we wish to measure the adjectives' biases toward. In our case \(S_{A}\) and \(S_{B}\) are representative sets of words for the concepts \textit{``masculinity''} and \textit{``femininity''} containing the words [``male'', ``man'', ``boy'', ``masculinity'', ``masculine'', ``dad'', ``father'', ``son''] and [``female'', ``woman'', ``girl'', ``femininity'', ``feminine'', ``mom'', ``mother'', ``daughter'']. These sets of words are obtained from the combination of suggested attribute words by Caliskan et al. and Ferrer et al. \cite{Caliskan2017, ferrer2021discovering}. Now let \(c_A\) and \(c_B\) be the weighted centroids of \(S_{A}\) and \(S_{B}\). We measure each adjective's relative bias strength towards \(S_{A}\) by the subtraction of its cosine similarity with \(c_A\) from its cosine similarity with \(c_B\) as in Equation \ref{equation:bias}:

\begin{equation}
    B_{w_i,S_A|S_B}=cos(w_i,c_A)-cos(w_i,c_B)
    \label{equation:bias}
\end{equation}

We apply this formula to all the adjectives present in each subreddit using Continuous Bag of Words (CBOW) as our word embedding algorithm; an unsupervised Deep NLP algorithm that is designed to predict a target word based on its context (surrounding words). Thus, ideally, words that appear in the same context tend to have higher cosine similarities.

Next to the words related to male identity and female identity, we also use two sets of attribute words consisting of male vs female pronouns [``he'', ``him'', ``his''] and [``she'', ``her'', ``hers''] in a parallel analysis to measure the toxicity toward male and female individuals rather than male and female identity. This will be explained in Section~\ref{sec:results} in detail.

\subsection{Sexism Indicator}
\label{subsec:sexismindicator}

To calculate an indicator that can quantify toxicity toward men and women in a community, we separate each community's adjectives' list into biased toward male attribute set (man, boy, father, etc.) vs biased toward female attribute set (woman, girl, mother, etc.). 
In parallel, we also separate the adjectives' lists into biased toward masculine pronouns attribute set (he, him, his) vs biased toward feminine pronouns attribute set (she, her, hers) using the same formula. 
Then, the toxicity targeted toward every attribute set is calculated by averaging  the three variables introduced in Section~\ref{subsec:preliminaries}.

Equation \ref{equation:TargetedToxicity} describes our model for calculating the toxicity towards attribute set \(S_A\) w.r.t. attribute set \(S_B\) assuming that \(w_i\) is in the set of adjectives that are biased towards \(S_A\): 

\begin{equation}
    TargetedToxicity_{S_A|S_B} = \frac{\sum_{i}\{B_{w_i,S_A|S_B} \times FPR_{w_i} \times T_{w_i}\}}{|\{w_i\}|}
    \label{equation:TargetedToxicity}
\end{equation}

\noindent Consequentially, if we replace \(S_A\) with the female attribute-set and \(S_B\) with the male attribute-set, the formula would give us a measurement of toxicity toward women. Swapping \(S_A\) and \(S_B\) would quantify the level of toxicity toward men in a community.

Moreover, we differentiate between toxicity toward female and male identity vs toxicity toward individual female or male figures. 
These two were often mixed in the previous literature associated with Word Embedding Association Tests. 
To enable our model to distinguish between the two types of targeting, we also apply attribute-sets of male vs female \emph{pronouns} on \(S_A\) and \(S_B\) (i.e. \([he, his, him]\) vs \([she, her, hers]\)) to get a measurement of toxicity targeted toward \emph{individual} male and female characters. 
This is critical in the sense that a community might target a certain group of male/female folks, yet harass them regardless of their gender. For instance, users of a political online community in a male-dominated country are likely to target their politicians, that may more likely be male figures, while not being a misandrist community. In that case, a word-embedding association test that mixes male pronouns (i.e. he, his, him) and male identity terms (e.g. man, masculinity, etc.) to form the attribute-set of \textit{men} might return impure results. 

For every subreddit (the online community type we analyze), we bootstrap 100,000 comments from the total data and repeat the experiment ten times with different seeds. Then, we calculate a confidence-interval for each subreddit based on the ten samples. The results for each attribute-set and subreddit are presented in Section~\ref{sec:results}. 

\section{Datasets}
\label{sec:datasets}


Our work uses two sources of data: 
1) a collection of comments from multiple subreddits, and 2) a set of annotated comments (toxic vs non-toxic). 
We leverage the former to measure the sexism rate of these communities, and the latter to build our supervised toxicity-detector NLP model. 
Next, we describe our data sources, starting with a brief introduction to the subreddits in our dataset and a characterization of each community according to the previous studies. 

\subsection{Subreddits}
\label{sec:datasets:Subreddits}

We query Reddit using the Pushshift API \cite{pushshift}. 
In particular, we query all comments from the following subreddits: \vspace{.1cm}

\noindent \textbf{r/TRP}, \textit{TheRedPill}, is a sub-movement of The Men’s Rights Activism (MRA) movement that offers advice to men regarding how to protect their masculinity that ``is under threat by the society'' \cite{Mountford2018}. It tends to ``empower'' heterosexual men with seduction strategies by exploiting arguments from evolutionary psychology \cite{Valkenburgh2021,papadamou2021over}. On the other hand, \textbf{r/MGTOW}, \textit{Men Going Their Own Way}, is another manosphere subreddit that encourages men to separate their path from women as a means of protection from a society ``corrupted by feminism'' \cite{Wright2020}. Previous literature has categorized both r/TRP and r/MGTOW as misogynist subreddits \cite{Dignam2019} and they were banned from Reddit in 2018 and 2021 respectively. \textbf{r/MGTOW2} also seems to be the continuation of the latter subreddit and was also banned from Reddit in 2021.\vspace{.1cm}

\noindent \textbf{r/FemaleDatingStrategy} defines itself as a ``female-exclusive subreddit that offers dating strategies for women who want to take control of their dating lives''. There are reports of the community's tendency to objectify the opposite gender and has been accused by r/AgainstHateSubreddits of encouraging transphobic and misandrist attitudes. Yet, there are also reports of using misogynist slang such as ``pickmeisha'' (a woman who lowers standards to receive attention from men) and ``cockholm syndrome'' (when a woman keeps going back to ``low-value'' men) \cite{Taylor2020}.\vspace{.1cm}

\noindent \textbf{r/IncelTear} defines itself as a community that tends to post screenshots of ``misogynistic'' and ``hateful'' comments from ``incels'' (involuntary celibates) in order to criticize them sarcastically. Next to the usage of irony, the community also has a record of re-posting and quoting extremely misogynist comments from the r/Incels community with the aim of sarcasm \cite{Dynel2020}. Ironically, r/IncelTear contains even more misogynist terms than r/MGTOW \cite{Farrel2019}, probably due to its high rate of quoting the most extreme misogynist comments from manosphere communities.\vspace{.1cm}

\noindent \textbf{r/TrollXChromosomes} is a subreddit designed for posting feminist humor and memes in order to criticize some aspects of the ``hegemonic femininity'' \cite{Massanari2017}. On the other hand, \textbf{r/TrollYChromosome} is known as a progressive subreddit for men casting sarcasm and humor toward the attitudes of society toward men and masculinity \cite{Myers2020}. \textbf{r/MensRights}, \textbf{r/MensLib}, \textbf{r/theGirlSurvivalGuide}, \textbf{r/Feminism}, \textbf{r/AskFeminists}, and \textbf{r/AskWomen} are the other subreddits from the gender-related discourse which we were interested in discovering their attitude to cover the whole spectrum of the discourse.

\subsection{Supervised Toxic Data}

To build and validate our sexism's indicator toxicity detector part (recall that the other parts of our sexism indicator are unsupervised, and hence do not need to be trained), 
we combine five different annotated toxic datasets from multiple sources to cover various sorts of toxicity:

\begin{itemize}
    \item OffensEval 2019\footnote{\url{https://sites.google.com/site/offensevalsharedtask/offenseval2019}} was one of the tasks in SemEval 2019\footnote{\url{https://alt.qcri.org/semeval2019/}} for detection of offensive language. It consisted of three sub-tasks. A: Identifying offensive language. B: Categorizing the offense. C: Identifying the Target of the offense \cite{Zampieri2019offenseval}. We ignore B and C and only consider the labels (\textit{offensive} 4400 or \textit{not-offensive} 8840) from task A. As explained in Section~\ref{sec:related-work}, our goal is to cover the C subtask using the effect of word-embedding associations.
    \item A dataset by Kaggle containing three labels (\textit{hate-speech} 1430, \textit{offensive} 19190, \textit{and} 4163) \cite{Samoshyn2020}. We joined \textit{hate-speech} and \textit{offensive} together as the \textit{Toxic} label and \textit{neither} as \textit{non-Toxic}.
    \item \textit{Wikipedia Talk Labels} dataset containing 100k discussion comments from the English Wikipedia \cite{Wulczyn2017}. Around 13k of them were labeled as personal attacks which we included in our \textit{Toxic} class and the rest in the \textit{non-Toxic} class.
    \item \textit{Toxic Comment Classification Challenge}\footnote{\url{https://www.kaggle.com/c/jigsaw-toxic-comment-classification-challenge}} dataset by Kaggle containing 15k \textit{Toxic} and 140k \textit{not-Toxic} annotated comments from Wikipedia.
    \item \textit{Jigsaw Multilingual Toxic Comment Classification}\footnote{\url{https://www.kaggle.com/c/jigsaw-multilingual-toxic-comment-classification/rules}} data containing 20k toxic and 200k not-toxic comments from Wikipedia. We only add the 20k toxic comments to our \textit{Toxic} class and leave the \textit{non-Toxic} ones for validation purpose (see Section~\ref{subsec:validationdatasets}).
\end{itemize}

After merging all the annotated data and splitting them into 70\% train-set and 30\% test-set, we obtain 52k comments for the \textit{Toxic} class and 180k for the \textit{non-Toxic} in the train-set.

\jose{I finally moved this here, so we have everything about supervised data together}
For evaluating the built supervised toxicity detector, 
we retain 30\% of the total supervised data for testing the F1-Score of our supervised \textit{Toxicity-Detector} NLP model.
In particular, we create a large dataset of non-Toxic comments by aggregating the non-toxic comments from the 30\% test-set and all the 200k non-toxic comments in \textit{Jigsaw Multilingual Toxic Comment Classification}. 
This makes it a total of 270k non-Toxic comments. \jose{is this the one we use in 5.1? if so, just say it from 5.1 as well to make it clear}\vahid{added in 5.1}

\subsection{Sexism Indicator Evaluation Datasets}
\label{subsec:validationdatasets}

For evaluating our final sexism indicator, we use two main datasets:

\begin{enumerate}
    \item 

We collect 1.2m random comments from the most recent comments on Reddit through the Pushshift API. 
These comments let us assess how our metric compares when run on a misogynist subreddit (cf. Section~\ref{sec:datasets:Subreddits}) and when run on a random dataset taken from all Reddit communities.

\item
We assemble an annotated dataset of ground-truth \emph{misogynistic comments} from three sources: \cite{guest-etal-2021-expert}, Kaggle\footnote{\url{https://www.kaggle.com/code/kerneler/starter-sexist-workplace-statements-a8e79cab-c/input}}, and \cite{chernyshev-etal-2023-lct} with a total of $\approx30000$ comments annotated as misogynistic or not ($\approx6000$ misogynistic), and use it as described in the next section.
\end{enumerate}




\section{Evaluation}
\label{sec:evaluation}
Before using our sexism indicator to analyse the subreddits stated in Section~\ref{sec:datasets:Subreddits}, the results and discussion of which we present later on in Section~\ref{sec:results}, we evaluate the reliability of our sexism metric. We do this in two steps. The first step is to evaluate the supervised ML part of our sexism metric, that is, the toxicity detector, as it is standard to do in supervised ML approaches to have confidence in the model trained (note that, obviously, we did not need to train any models for the \emph{unsupervised} parts of the metric, i.e., the frequency ranking and the bias embedded). The second step is the evaluation of our sexism metric as a whole, particularly showing that it is sensitive and it increases its value as more sexist comments are added. 

\subsection{Evaluation of the supervised toxicity detector} 
The Toxicity-Detector NLP model of our sexism metric was evaluated by splitting the dataset in Section~\ref{subsec:validationdatasets} into 70\% train-set and 30\% test-set\jose{todo - refer to 4.2?}\vahid{added}. Table~\ref{tab:conf-td} shows the confusion matrix we obtained on the test-set. Our F1-Score macro was above 91\% and precision and recall macro were above 90\% and 92\% respectively.
\begin{table}[h]
    \centering

    \begin{tabular}{cc|cc|}
    \multicolumn{2}{c}{}
            &   \multicolumn{2}{c}{Predicted} \\
    &   \textit{label} &   Toxic &   Not-Toxic        \\ 
    \cline{2-4}
    \multirow{2}{*}{\rotatebox[origin=c]{90}{Actual}}
    & Toxic   & 20346 & 2331           \\
    & Not-Toxic    & 4085 & 73691          \\ 
    \cline{2-4}
    \end{tabular}    
    \caption{Confusion Matrix for Toxicity-Detector model}
    \label{tab:conf-td}
\end{table}

We also examined a well-known pre-trained transformer-based language model for toxicity-detection called Toxic-BERT\footnote{\url{https://huggingface.co/unitary/toxic-bert}} which yielded a slightly better performance on the same test-set (93\% F1-Score macro). However, this model and similar large neural networks maintain a significantly higher computational cost in comparison to classical models. 
In particular, our classical NLP model takes 10 seconds to machine-annotate 100k samples, whereas the same tasks take more than 1 hour for Toxic-BERT. 
As our final holistic model and research question deal with huge corpora, we prefer to stick to a computationally cheaper model where not much accuracy is sacrificed.

\subsection{Evaluation of the Sexism Metric}
\begin{figure}[t]
  \centering
  \includegraphics[width=.99\linewidth]{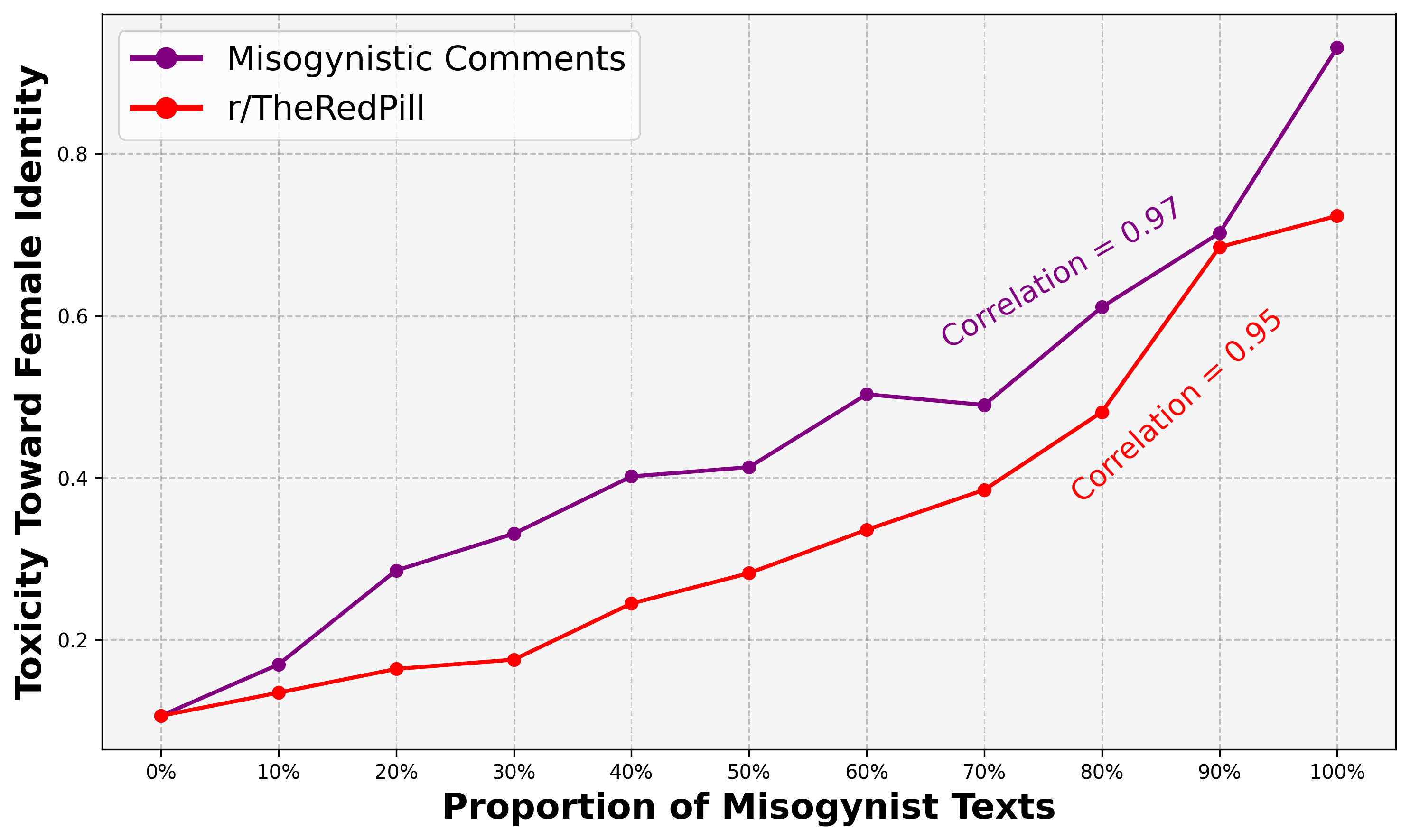}
  \caption{Validation Chart for Our Sexism Metric for Toxicity Toward Female Identity\jose{Vahid - you need to change the title of the Y axis to make it consistent with the caption}\vahid{I changed the caption instead}}
  \label{fig:misogyny-validation}
\end{figure}

The second aspect that we evaluate is the reliability of our sexism metric. We iteratively create 10 different datasets, each with a different level of misogynistic comments, and observe if our metric increases as we increase the amount of misogynistic comments in the data. Our data-generation strategy is to start with a dataset composed of neutral comments and gradually increment the misogyny of the dataset by adding subsets of misogynistic comments. We take what's left of the supervised training (comments that fall out of our train-set) from the non-Toxic class as a starting point. Then, we use external $\approx6000$ comments annotated as ``misogynist'' as our pool of sexism toward female identity. Moreover, since there is a consensus in the previous literature in labeling r/TRP as a ``misogynist'' community (e.g. \cite{Dignam2019, Ging2019, Farrel2019}), we consider r/TRP comments as our ground-truth for another pool of misogynist data to be analyzed separately. %
Then, for each pair of ``misogynist comments vs. neutral'' and ``r/TRP vs. neutral'' data we iteratively form ten new datasets with different bootstrapped proportions of the misogynist pool and run our misogyny-detection formula from Section~\ref{equation:TargetedToxicity} on each of the ten datasets (per pair) to get ten different values of our sexism metric toward female identity.

We observe an above 97\% correlation between the proportion of annotated misogynistic comments and our sexism metric value towards female identity. This correlation is also above 95\% for the scenario where we use different proportions of r/TheRedPill comments vs. neutral ones. 
The score for each iteration in both cases is illustrated in Figure \ref{fig:misogyny-validation}. Note that our evaluation can only be limited to the case of toxicity toward female identity since misandry is an understudied concept, and there is none or insufficient ground-truth of misandrist or toxic-toward-female-individuals datasets available online.

\jose{should the following paragraph not be moved to the next section? we are talking about results that we have not yet talked about? - it could be the last of the bold-headed paragraphs and it could be called "Comparison with other approaches"??}\vahid{yeap. unfortunately, this is another side effect of putting the evaluation section before the results. what I have been arguing not to do.}

\section{Results \& Discussion} 
\label{sec:results}

\begin{figure*}[h]
  \centering
    \begin{minipage}[b]{0.045\linewidth}
      \centering
      \includegraphics[width=1.00\linewidth]{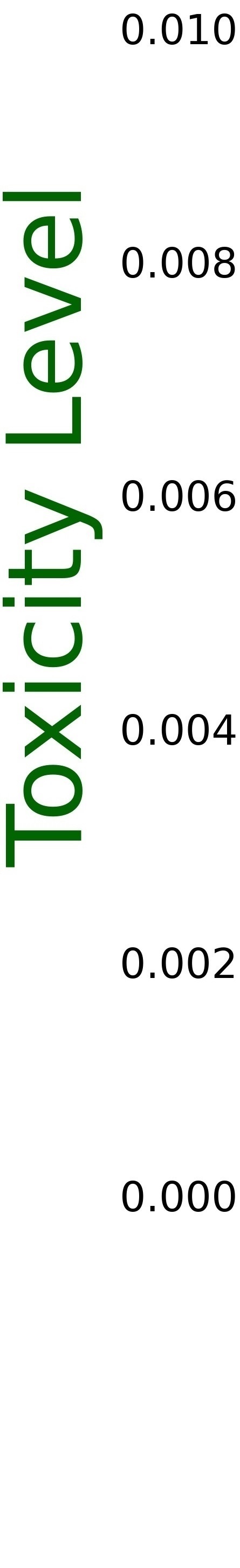}
  \end{minipage}
  \begin{minipage}[b]{0.219\linewidth}
      \centering
      \includegraphics[width=\linewidth]{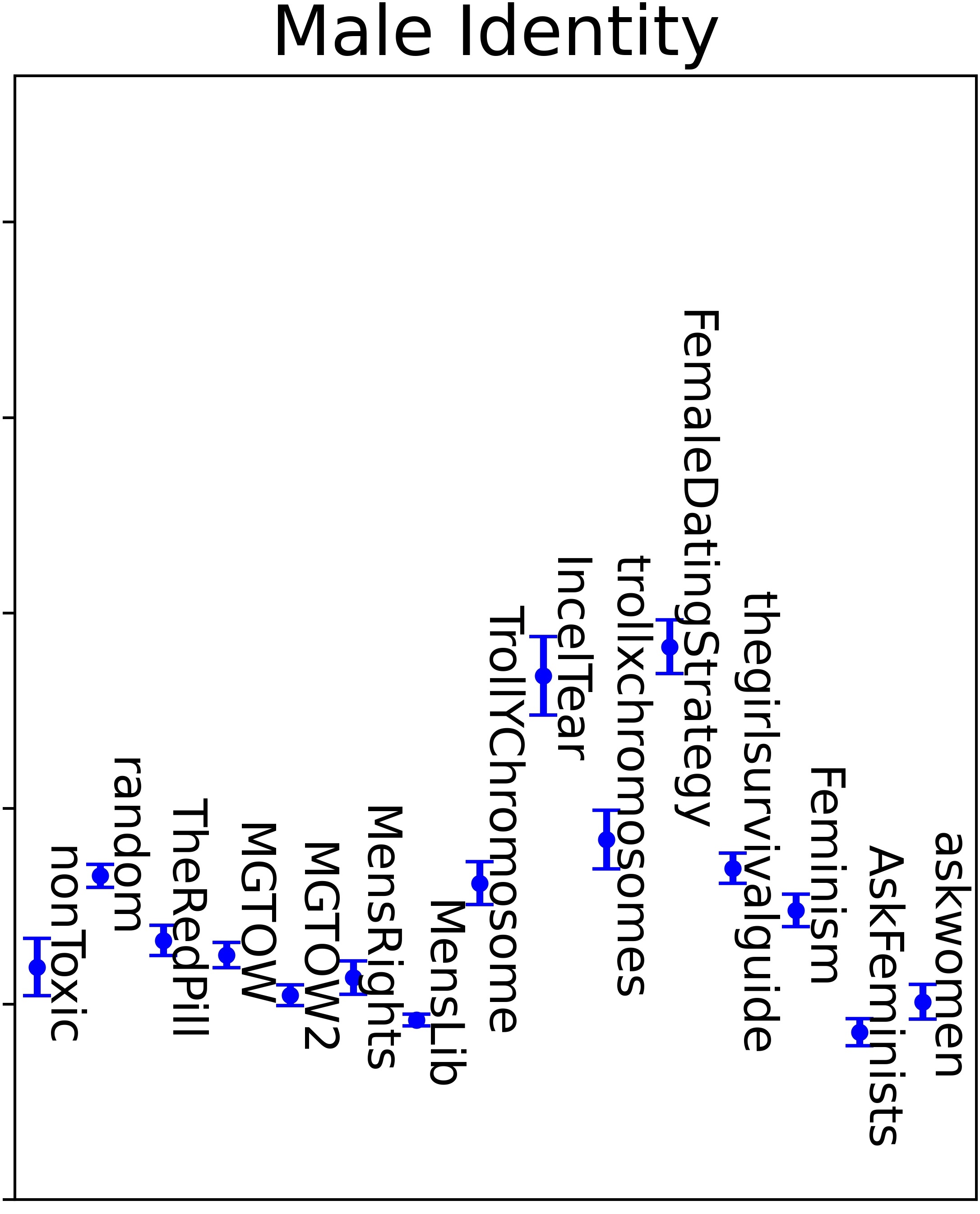}
      \caption{Toxicity Targeted Toward Male Identity.}
      \label{fig:misandry-ethnic}
  \end{minipage}
  \hfill
  \begin{minipage}[b]{0.219\linewidth}
      \centering
      \includegraphics[width=\linewidth]{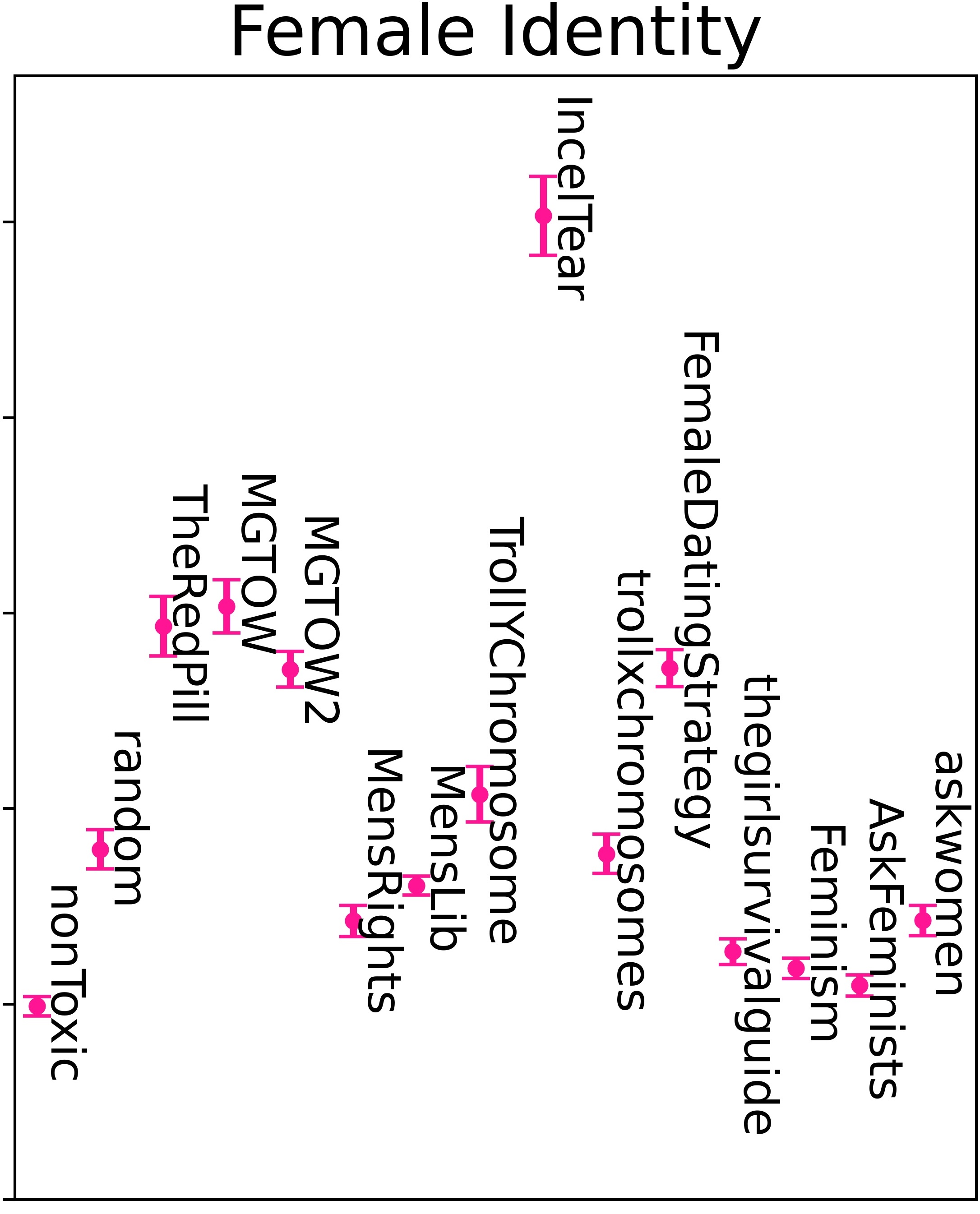}
      \caption{Toxicity Targeted Toward Female Identity.}
      \label{fig:misogyny-ethnic}
  \end{minipage}
  \hfill
  \begin{minipage}[b]{0.219\linewidth}
      \centering
      \includegraphics[width=\linewidth]{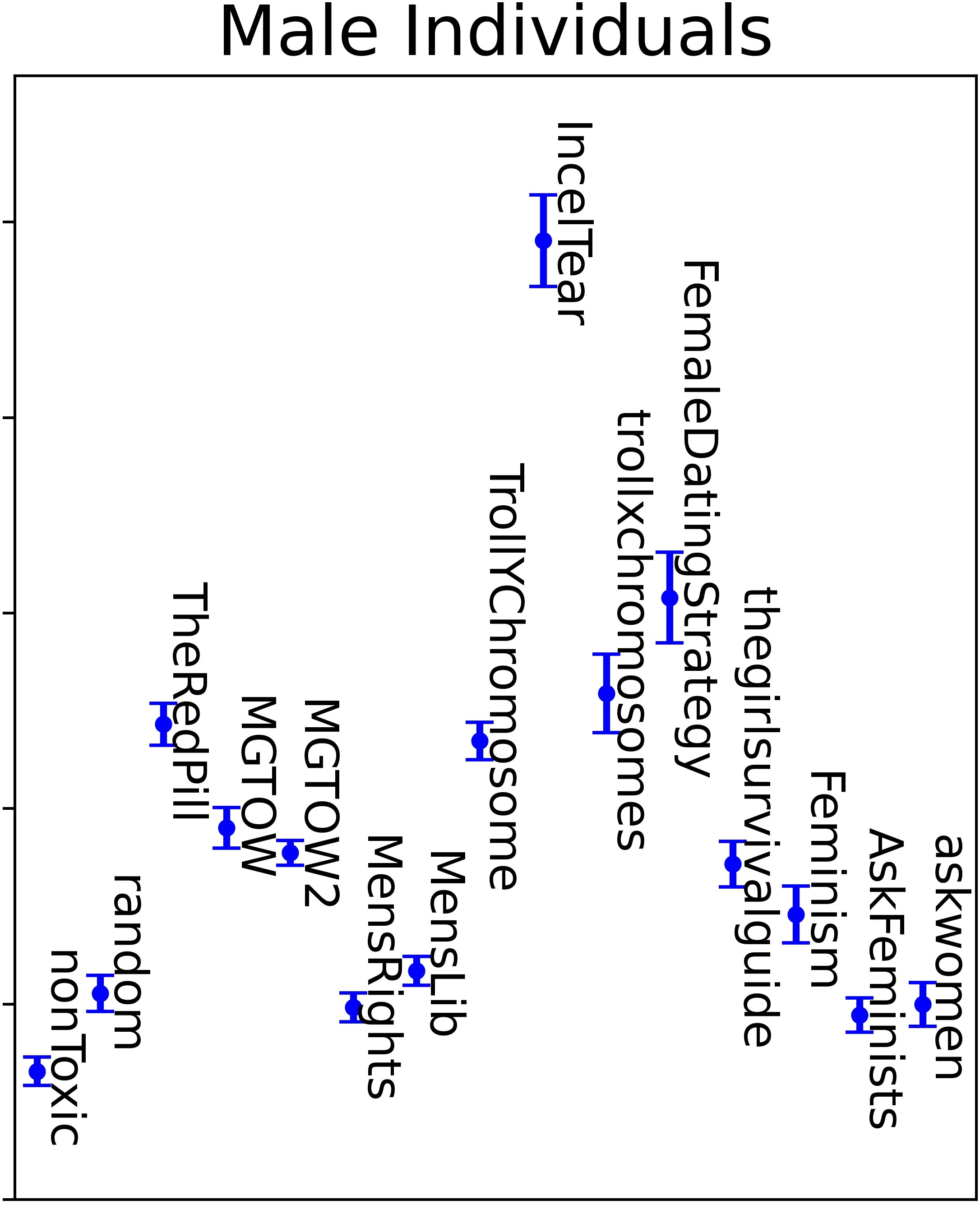}
      \caption{Toxicity Targeted Toward Male Individuals.}
      \label{fig:misandry-individual}
  \end{minipage}
  \hfill
  \begin{minipage}[b]{0.219\linewidth}
      \centering
      \includegraphics[width=\linewidth]{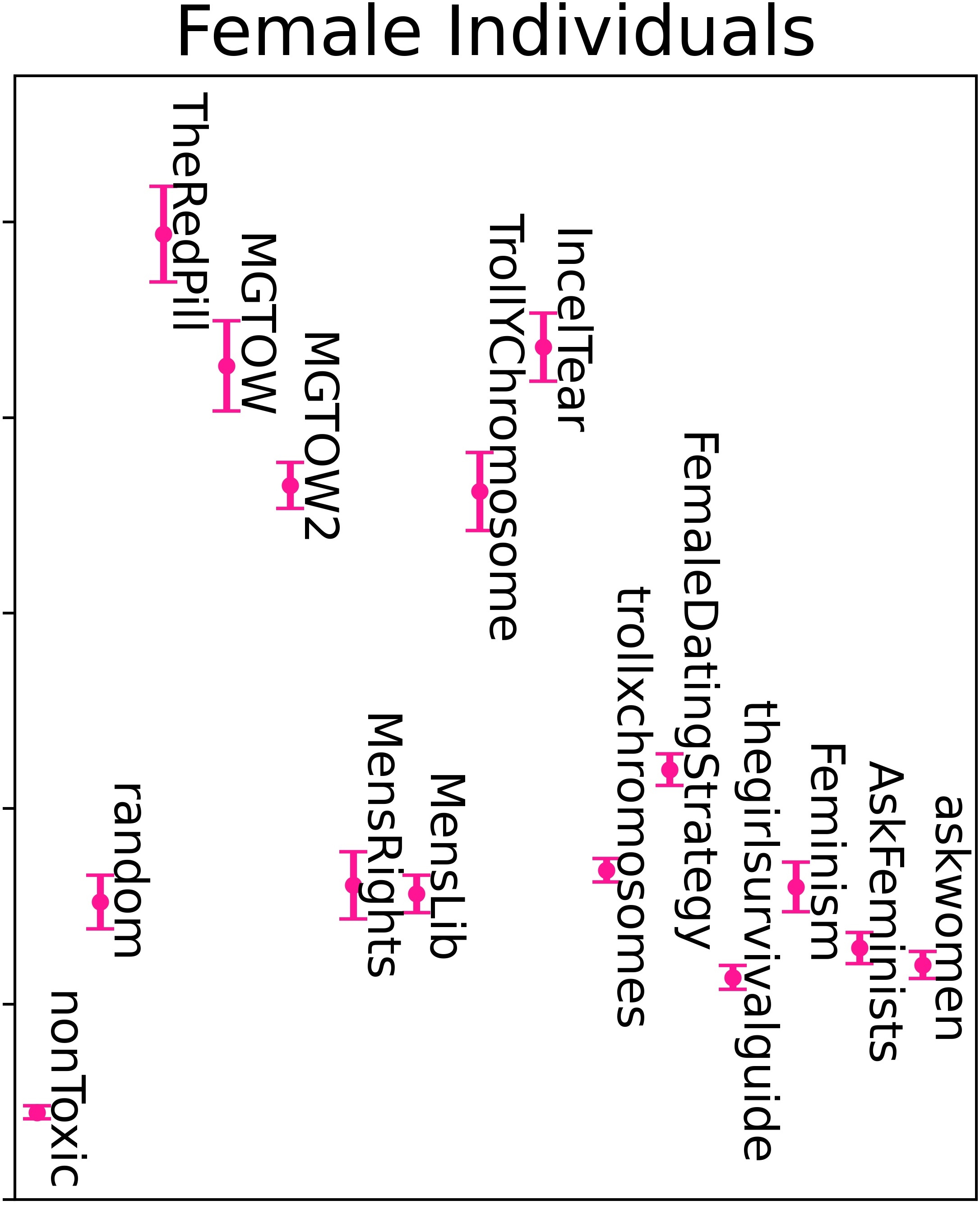}
      \caption{Toxicity Targeted Toward Female Individuals.}
      \label{fig:misogyny-individual}
  \end{minipage}
\end{figure*}

\noindent{\bf Toxicity towards identity.}
Figures \ref{fig:misandry-ethnic} and \ref{fig:misogyny-ethnic} show the \textit{toxicity towards male identity} and \textit{toxicity towards female identity} in each of the subreddits that can be interpreted as the level of misandry and misogyny inside each of them. The level of toxicity for each subreddit is obtained through Equation \ref{equation:TargetedToxicity} separately for adjectives biased toward male identity and adjectives biased toward female identity.

The vertical error-bars for each subreddit shows the 95\% confidence interval of the metric after 10 bootstraps. The non-Toxic corpus on the left acts as the baseline point of the metric for a non-toxic community. The second corpus from the left, also shows the level of targeted toxicity for a randomly collected set of comments from Reddit to assess which targeted toxicity is more salient than usual Reddit discourse.

\noindent{\bf Toxicity towards individuals.}
On the other hand, Figures \ref{fig:misandry-individual} and \ref{fig:misogyny-individual} refer to the toxicity targeted towards male and female pronouns attribute-sets in each of the subreddits that can be interpreted as the level of toxicity targeted toward individual male and female characters. There are significant visible changes with respect to targeted toxicity toward gender identity. For instance, r/IncelTear proves highly toxic toward individual male figures rather than the male identity. The same case happens for manosphere communities (r/TheRedPill, r/MGTOW, and r/MGTOW2) with respect to toxicity toward female individuals.

\noindent{\bf Findings.} 
The results detect various levels of targeted toxicity with the highest targeted toward female individuals next to some cases of targeted toxicity toward men. 
The targeted toxicity index we obtain for every community confirms existing analyses looking at the toxicity of r/TRP, r/MGTOW, r/MGTOW2, r/IncelTear, and r/FemaleDatingStrategy; all the subreddits in which we had a prior report of their toxicity in the literature (more information in Section~\ref{sec:evaluation}). 

In addition, {\em our framework helps uncover the characteristics of subreddits that have not been analyzed before}. 
Consistent with the descriptions mentioned in Section~\ref{sec:datasets:Subreddits}, r/TRP, r/MGTOW, and r/MGTOW2 stand among the top scores in toxicity both toward female identity and towards female individuals yet they are more extreme with respect to female individuals rather than female identity.
\begin{itemize}
    \item r/FemaleDatingStrategy acquired a significant score of toxicity both towards men and women; consistent with a qualitative report \cite{Taylor2020} described in Section~\ref{sec:datasets:Subreddits}. 
    
    \item r/IncelTear was seen as high in terms of toxicity toward both men and women. Its excessive toxicity toward women, in terms of both identity and individual, is not counter-intuitive due to the nature of the subreddit causing it to repeatedly and sarcastically narrate misogynistic comments from ``incels'' prior to humiliating them. Moreover, the community shows the most salient score in toxicity toward \textit{individual} male figures, which perfectly makes sense looking at the agenda of the subreddit --- that is, to target individual male ``incels'' regarding their ``misogynist'' comments. r/TrollYChromosome and r/TrollXChromosomes obtained almost a symmetric score regarding the gender they cast their toxic content toward as if they were twins from both sides of the spectrum. In addition, they seemed to be respectively more toxic toward female and male individuals rather than female and male identities.

    \item r/MensRights and r/MensLib were more toxic toward women than men. 
    However, their level of toxicity did not significantly exceed a dataset of random comments from Reddit. This can presumably be attributed to more recent moderation policies imposed by Reddit and communities' moderators and their difference of ideology from right-wing MRA movements (e.g. r/TheRedPill, MGTOW). We know that despite some other MRA subreddits, these two subreddits have in fact not been banned by Reddit till now, which is compatible with the results of our indicator.
\end{itemize}

\begin{table}[h]
  \centering
  \caption{Top-100 most salient terms similarity matrix. The top-right (red) side of the table shows the number of common adjectives among the top 100 saliently biased adjectives toward female-identity. The bottom-left (blue) side depicts the same quality for male-identity.}
  \includegraphics[width=\linewidth]{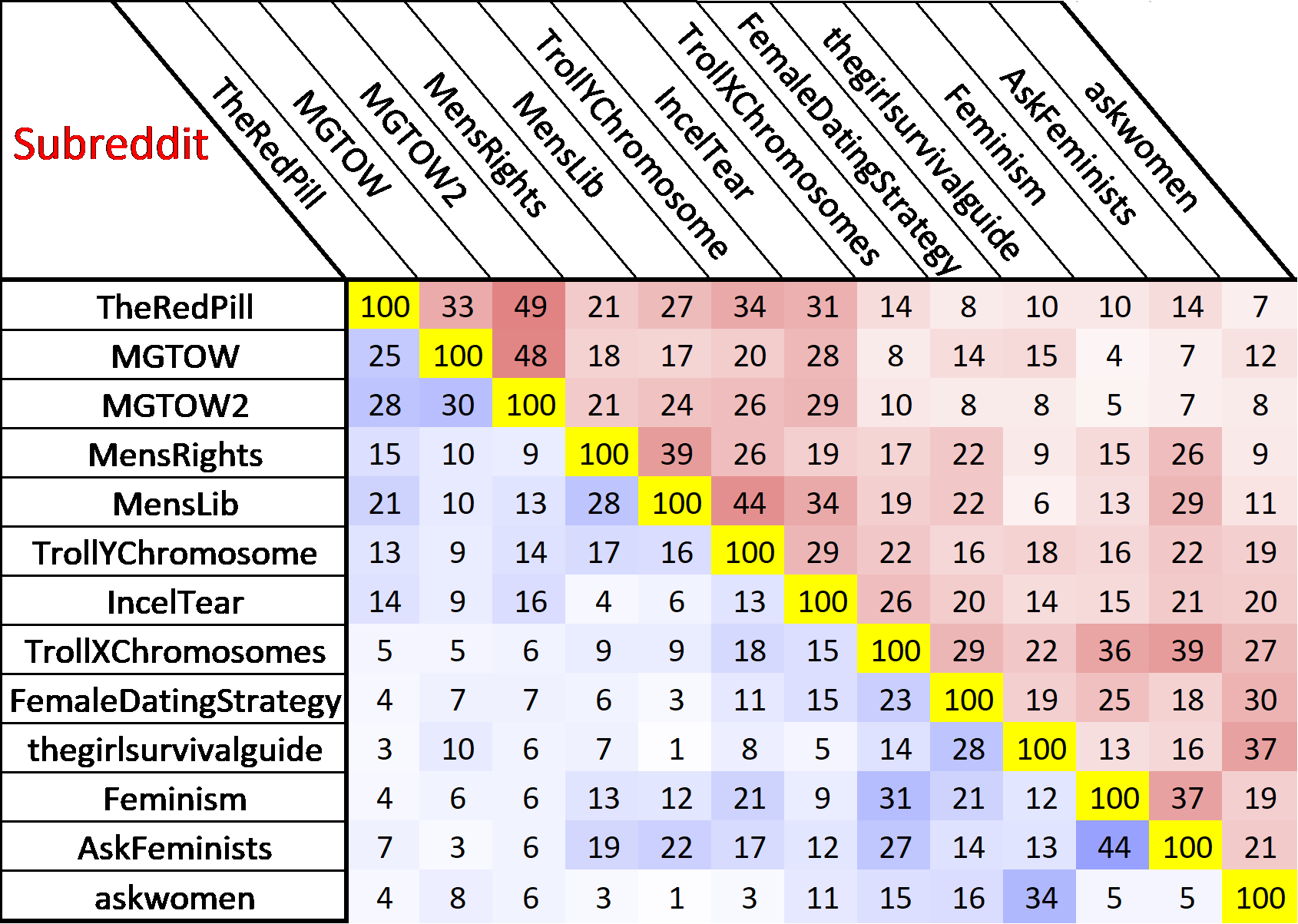}
  \label{tab:similarity-matrix}
\end{table}
\noindent{\bf Saliency.}
As a further comparison between gender-discourse among the subreddits, we create a similarity index for every pair of them. We use the notion of \(saliency=B_{w_i,S_A|S_B} \times FPR_{w_i}\) \cite{ferrer2021discovering} to sort the adjectives inside every subreddit based on how biased-\&-frequent (salient) a word is toward female and male identity. Then, in Table \ref{tab:similarity-matrix} we quantify this gender discourse similarity. 
The top-right side of Table \ref{tab:similarity-matrix} (red cells) shows the number of top-100 most saliently biased adjectives toward female-identity that are common among a pair of subreddits. 
The bottom-left side of the table (blue cells) shows the same concept for top-100 most saliently biased adjectives toward male-identity.

The similarity rates share several insights regarding how far or close discourses are for each pair of subreddits:

\vspace{.25cm}
\noindent{\em Insight \#1.} 
r/TheRedPill, r/MGTOW, and r/MGTOW2 contain a highly similar discourse both in male-related and female-related adjectives among each other, while being more similar in the latter. This can show that manosphere subreddits agree more on what to attribute to femininity than to masculinity. 

\vspace{.25cm}
\noindent{\em Insight \#2.} 
r/MensRights and r/MensLib have few common words with r/TheRedPill, r/MGTOW, and r/MGTOW2 about women, which would make sense as our model also has not rated them as highly toxic toward women. Therefore, they have less similar discourse to subreddits that are conventionally recognized as misogynistic by the previous literature. 

\vspace{.25cm}
\noindent{\em Insight \#3.} 
r/MensRights and r/MensLib are mostly similar to each other, especially in their share of female-related adjectives. Interestingly, in comparison to manosphere subreddits, they have higher rates of similarity with feminist subreddits in both their top male-associated (e.g., \textit{``paternalistic''}, or \textit{``misogynistic''}) and female-associated adjectives (e.g., \textit{``successful''}, or \textit{``powerful''}). 
\citeauthor{LaViolette2019} raise discussions on how r/MensRights and r/MensLib share highly similar lexical features to talk about the same topic from an anti- and pro-feminist perspective \cite{LaViolette2019}.

\vspace{.25cm}
\noindent{\em Insight \#4.} 
r/IncelTear tends to show high discourse-similarity with r/TheRedPill, r/MGTOW, and r/MGTOW2 regarding women (e.g., \textit{``promiscuous''}, \textit{``hypergamous''}, or \textit{``casual''}), but not much similarity regarding men. 
This is compatible with the previous descriptions of r/IncelTear as a community that tends to sarcastically \emph{quote} misogynist comments and humiliate those comments. Those quotations of misogynist comments could be the probable reason behind the high amount of similarity in feminine-biased words with the manosphere subreddit. Yet, the lower commonality in words saliently biased towards men, reconfirms that they do not actually share the discourse of manosphere subreddits regarding masculinity.

\vspace{.25cm}
\noindent{\em Insight \#5.} 
Both feminist communities, r/Feminism, and r/Ask\-Fe\-mi\-nists, expectedly show high discourse similarity among themselves regarding both men (e.g., \textit{``arrogant''}, or \textit{``misogynistic''}) and women (e.g., \textit{``independent''}, or \textit{``successful''}). 
Moreover, they are considered similar to r/trollxchromosomes which is known as a subreddit meant for feminist humor \cite{Massanari2017}. r/TheGirlSurvivalGuide and r/FemaleDatingStrategy, as two daily-life and dating tips subreddits, share the highest similarity in their male-related terms such as \textit{``unemployed''}, or \textit{``unsuccessful''}, etc. which might be attributed to the types men suggested to avoid in dating.

\vspace{.25cm}
\noindent{\em Comparison with Fully Supervised Approaches:}

Finally, we also compare the results we obtained with previous fully supervised approaches for overall toxicity of corpora~\cite{Farrel2019,Ribeiro2020}. For this, we compare our results in the particular subreddits used in those previous works, which were consistent with our results. Like us, \citeauthor{Farrel2019} rated r/IncelTear as more misogynistic comments than r/MGTOW, and \citeauthor{Ribeiro2020} rated r/MGTOW and r/TRP as almost equally toxic. They also measured a random set of Reddit comments to have slightly more than half the toxicity level of r/TRP \cite{Farrel2019,Ribeiro2020}.  
%
Yet, both of the mentioned works are limited to fully-supervised methods, simply counting the percentage of misogynistic posts and context-unaware lexicons inside a community. This will make them expensive to annotate, more dependent on subjective judgments (i.e., whether a post should be annotated as misogynist or not), less robust, and less generalizable to other sorts of targeted toxicity (e.g., for measuring targeted toxicity toward Muslims vs non-Muslims, they should run a separated analysis with toxic jargons related to Islamophobia).
It will also make them unable to detect neutral-looking lexicons that could be toxic in certain communities' contexts and vice versa.

\section{Conclusion \& Future Works}
\label{sec:discussion}

In this paper, we proposed a metric that is able to \emph{quantify} the level of sexism in the language of online communities, using a combination of unsupervised and supervised NLP techniques. Our analysis embraced 14 subreddits from different parts of the gender-discourse spectrum, which were not analyzed before by a unique model at the same time. We confirmed the toxicity of r/TheRedPill and r/MGTOW toward women in an automated and comparable way. 
We also realized that a female-exclusive dating community such as r/FemaleDatingStrategy can be toxic toward men and women at the same time.
The granularity of our method to distinguish the target of toxicity offers a new nuanced understanding of Web communities, which will foster future work in the area. 

Likewise, another contribution of our method and subsequent analyses was making a clear distinction between the toxicity aimed toward male/female identity inside a community, and the toxicity targeting male/female individuals. 
This enables better attribution mechanisms, which is paramount to curve misinterpretations about a community when there is abundant criticism toward several male/female politicians rather than its toxic content about male/female ``identity''. 



Furthermore, our model can smoothly be generalized to capture other sorts of polarization and radicalization on social media. For instance, by changing our attribute words with sets of words related to Democrats and Republicans, and replacing our Embedded-Toxicity parameter with Embedded-Polarity, one can be able to scalably quantify the polarization of \emph{sentiments} towards the \emph{Democratic} and \emph{Republican} parties in different timelines. 

One feature of our methodology is that it accepts any type of embedded biases, which opens new avenues to offer more granularity to the identification of toxicity. 
For instance, another implementation for future work could detect the polarization of sarcasm or humor targeted toward either group, by simply replacing the Embedded-Toxicity parameter with Embedded-Sarcasm. The facilitation provided by our approach would be that the researcher does not need to annotate two sparse binary classes of \textit{sentiment/sarcasm toward Republicans} vs \textit{sentiment/sarcasm toward Democrats}. An already available dataset of sentiment, sarcasm, etc. would suffice and the model itself would detect the target while suffering less from annotators' subjective judgments and biases.

Our holistic indicator of polarization provides the tool for policymakers and moderators to take action about a community (e.g., subreddit) by inspecting the polarization level over time. 
Also, it can be used in computational social science research for measuring polarization over time, and causal inference between temporal polarization and various real-world events (e.g., elections, wars, COVID-19). However, when it comes to moderation, this holistic indicator should not be projected into individual comments in a community and cause moderators to treat every comment in a polarized community as a polarized comment. Judging individual comments of users require a higher level of supervision and care.
Also it is worth noting that our work is only analyzing male and female genders for now and analyzing the LGBTQ+ groups is out of the scope of this paper. Future work need to extend the model from a binary polarization detector to a more complex multidimensional association problem to address this limitation. 

We make our code and datasets available on GitHub to the researchers for reproduction and further developments\footnote{\url{https://anonymous.4open.science/r/Polarization-Indicator-6243}}. 

\bibliography{bib1}





\end{document}